\documentclass[ste,twoside]{stefano}

\usepackage{amsmath}
\usepackage{amssymb}
\usepackage{graphicx}
\usepackage{epsfig}
\usepackage{rotating}
\usepackage{cite}
\usepackage{color}

\def\makeheadbox{{
\hbox to0pt{\vbox{\baselineskip=10dd\hrule\hbox
to\hsize{\vrule\kern3pt\vbox{\kern3pt \hbox{  {\sc Physical Review
D} {\bf 71}, 076008-12 (2005) } \hbox{
{\sc{\color{blue}{dma}}[{\color{black}{imecc}}]{\color{red}{UniCamp}}
} \hspace*{10.3cm} {\color{blue}{$\boldsymbol{\Sigma \delta
\Lambda}$}} } \kern3pt}\hfil\kern3pt\vrule}\hrule} \hss}}}

\def\0{\mbox{\tiny $0$}}
\def\1{\mbox{\tiny $1$}}
\def\2{\mbox{\tiny $2$}}
\def\3{\mbox{\tiny $3$}}
\def\4{\mbox{\tiny $4$}}
\def\5{\mbox{\tiny $5$}}
\def\6{\mbox{\tiny $6$}}
\def\7{\mbox{\tiny $7$}}
\def\8{\mbox{\tiny $8$}}
\def\9{\mbox{\tiny $9$}}
\def\a{\mbox{\tiny $\alpha$}}
\def\b{\mbox{\tiny $\beta$}}

\def\scalar{\mbox{\tiny $scalar$}}
\def\Dirac{\mbox{\tiny $Dirac$}}

\def\f14{\mbox{\tiny $\frac{1}{4}$}}
\def\infm{\mbox{\tiny $-\infty$}}
\def\infp{\mbox{\tiny $+\infty$}}
\def\L{\mbox{\tiny $L$}}
\def\i{\mbox{\tiny $i$}}
\def\z{\mbox{\tiny $z$}}

\def\j{\mbox{\tiny $j$}}

\def\mi{\mbox{\tiny $-$}}
\def\pl{\mbox{\tiny $+$}}
\def\ppm{\mbox{\tiny $\pm$}}

\def\EPJC#1{{\em Eur. Phys. J.}\; {\bf C #1}}

\def\APB#1{{\em Acta Phys. Polon.}\; {\bf B #1}}

\def\PLB#1{{\em Phys. Lett.}\; {\bf B #1}}
\def\IJMP#1{{\em Int. J. Mod. Phys.}\; {\bf #1}}

\def\PRD#1{{\em Phys. Rev.}\; {\bf D #1}}

\def\PRP#1{{\em Phys. Rep.}\; {\bf #1}}

\def\PTP#1{{\em Prog. Theor. Phys.}\; {\bf #1}}

\def\AP#1{{\em Ann. Phys.}\; {\bf #1}}

\begin{document}
%

\title{FLAVOR AND CHIRAL OSCILLATIONS WITH DIRAC WAVE PACKETS}

\author{
Alex E. Bernardini\inst{1}
\and Stefano De Leo\inst{2}
}

\institute{
Department of Cosmic Rays and Chronology, State University of Campinas,\\
PO Box 6165, SP 13083-970, Campinas, Brazil,\\
{\em alexeb@ifi.unicamp.br} \and
Department of Applied Mathematics, State University of Campinas,\\
PO Box 6065, SP 13083-970, Campinas, Brazil,\\
{\em deleo@ime.unicamp.br}
}


\date{{\em January, 2005}}

\abstract{
We report about recent results on Dirac wave packets in
the treatment of neutrino flavor oscillation
where the initial localization of a spinor state
implies an interference between positive and negative energy
components of mass-eigenstate wave packets.
A satisfactory description of {\em fermionic} particles
requires the use of the Dirac equation as evolution equation
for the mass-eigenstates.
In this context, a new flavor conversion formula can be
obtained when the effects of chiral oscillation are taken into account.
Our study leads to the conclusion that the {\em fermionic} nature of the particles, where
chiral oscillations and the
interference between positive and negative frequency components of
mass-eigenstate wave packets are implicitly assumed, modifies the standard oscillation
probability.
Nevertheless, for ultra-relativistic particles and sharply peaked momentum
distributions, we can analytically demonstrate that these modifications introduce correction factors
proportional to $m_{\1,\2}^{\2}/p_{\0}^{\2}$ which are practically
un-detectable by any experimental analysis.
}










\titlerunning{Flavor and Chiral Oscillations with Dirac Spinors}

\maketitle

\section*{I. Introduction}

The Dirac formalism is useful and
essential in keeping clear many of the conceptual aspects of
quantum oscillation phenomena that naturally arise in a
relativistic spin one-half particle theory.
The quantum oscillation phenomena has stimulated
the analysis of several theoretical approaches \cite{Beu03,Bla03} on the flavor
conversion formula which, sometimes, deserve a special attention
because of carrying valuable physical information.
The applicability of the standard plane wave
treatment of oscillations by resorting to {\em intermediate}
\cite{Giu98,Zra98} and {\em external} \cite{Ric93,Giu02B} wave
packet frameworks has been extensively questioned in the last years \cite{DeL04,Beu03}.
Although the standard oscillation formula  \cite{Kay89,Kay04} could give the correct
result when {\em properly} interpreted,
the plane wave approach implies a perfectly well-known energy-momentum and an infinite
uncertainty on the space-time localization of the oscillating particle
which leads to the destruction of the oscillating character \cite{Kay81}.
The {\em intermediate} wave packet approach eliminates the most controversial
points rising up with the plane wave formalism.
Wave packets describing propagating mass-eigenstates
guarantees the existence of a coherence length \cite{Kay81}, avoids the ambiguous approximations in the plane
wave derivation of the phase difference \cite{DeL04} and, under particular conditions of
minimal {\em slippage}, recovers the oscillation probability given by the {\em standard}
plane wave treatment \cite{Ber04}.
Otherwise, a common argument
against the {\em intermediate} wave packet formalism is that oscillating
neutrinos are neither prepared nor observed \cite{Beu03}.
Some authors suggest the calculation of a transition probability between
the observable particles involved in the production and detection
process in the so-called {\em external}
wave packet approach \cite{Giu02B,Beu03}: the oscillating
particle, described as an internal line of a Feynman diagram by a
relativistic mixed scalar propagator, propagates between the
source and target (external) particles represented by wave
packets.
Anyway, it can be demonstrated \cite{Beu03} that the overlap function of
the incoming and outgoing wave packets
in the {\em external} wave packet model is mathematically equivalent
to the wave function of the propagating mass-eigenstate in the
{\em intermediate} wave packet formalism.
However, the overlap function takes into account not only the
properties of the source, but also of the detector. This is
unusual for a wave packet interpretation and not satisfying for
causality \cite{Beu03}. This point was clarified by Giunti
\cite{Giu02B} who solves this problem by proposing an improved
version of the {\em intermediate} wave packet model where the wave
packet of the oscillating particle is explicitly computed with
field-theoretical methods in terms of {\em external} wave packets.
In order concentrate the discussion
on the Dirac equation properties that we intend
to report in this manuscript, in this preliminary investigation,
we avoid the field theoretical methods in detriment to a clearer
treatment with {\em intermediate} wave packets
which commonly simplifies the
understanding of physical aspects going with the oscillation
phenomena \cite{DeL04,Tak01}.

Our final aim is the investigation of how the inclusion
of chiral oscillation effects can modify
the flavor conversion probability formula
which was previously obtained by using {\em fermionic} instead of {\em scalar}
particles, i. e. in treating the time evolution of the spinorial
mass-eigenstate wave packets, we shall take into account
the chiral nature of charged weak currents and the time evolution
of the chiral operator with Dirac wave packets.
To do it, we shall use the Dirac equation as the
evolution equation for the mass-eigenstates.
Before introducing
the Dirac formalism, in section II we briefly review the analytic calculations \cite{Ber04}
with the {\em intermediate} wave packet model for scalar particles \cite{Kay81}.
In particular, a {\em gaussian} wave packet is chosen to
describe the localization of the initial flavor state and to obtain
an analytical expression for the flavor conversion probability.
In section III, we shall recapitulate the Dirac
formalism \cite{Zub80, Ber04B} and show that a superposition of both positive and
negative frequency solutions of the Dirac equation is often a
necessary condition to correctly describe the time evolution of
mass-eigenstate wave packets.
The small modifications obtained in the context of a wave packet treatment of
oscillation phenomena are (briefly) compared with quantum field
theoretical calculations \cite{Beu03,Giu93,Bla95}.
In section IV, we notice that the use of Dirac equation solutions allows us to observe
the additional effect of chiral oscillation
already introduced by De Leo and Rotelli \cite{DeL98}.
As a natural extension, we show how to couple
chiral to flavor oscillations in the {\em intermediate} wave packet
framework. Finally, we give, for strictly
peaked momentum distributions and ultra-relativistic particles, an
analytic expression for the coupled flavor and chiral conversion probability.
We draw our conclusions in Section V.


\section*{II. Scalar Oscillating Particles}

The main aspects of oscillation phenomena can be understood by
studying the two flavor problem. In addition, substantial
mathematical simplifications result from the assumption that the
space dependence of wave functions is one-dimensional ($z$-axis).
Therefore, we shall use these simplifications to calculate the
oscillation probabilities. In this context, the time evolution of
flavor wave packets can be described by
\begin{eqnarray}
\Phi(z,t) &=& \phi_{\1}(z,t)\cos{\theta}\,\mbox{\boldmath$\nu_{\1}$} + \phi_{\2}(z,t)\sin{\theta}\,\mbox{\boldmath$\nu_{\2}$}\nonumber\\
          &=& \left[\phi_{\1}(z,t)\cos^{\2}{\theta} + \phi_{\2}(z,t)\sin^{\2}{\theta}\right]\,\mbox{\boldmath$\nu_\alpha$} +
          \left[\phi_{\1}(z,t) - \phi_{\2}(z,t)\right]\cos{\theta}\sin{\theta}\,\mbox{\boldmath$\nu_\beta$}\nonumber\\
          &=& \phi_{\alpha}(z,t;\theta)\,\mbox{\boldmath$\nu_\alpha$} + \phi_{\beta}(z,t;\theta)\,\mbox{\boldmath$\nu_\beta$},
\label{0}
\end{eqnarray}
where {\boldmath$\nu_\alpha$} and {\boldmath$\nu_\beta$} are flavor-eigenstates
and {\boldmath$\nu_{\1}$} and {\boldmath$\nu_{\2}$} are mass-eigenstates.
The probability of finding a flavor state $\mbox{\boldmath$\nu_\beta$}$ at
the instant $t$ is equal to the integrated squared modulus of the
$\mbox{\boldmath$\nu_\beta$}$ coefficient
\begin{eqnarray}
P(\mbox{\boldmath$\nu_\alpha$}\rightarrow\mbox{\boldmath$\nu_\beta$};t)
& = &
\int_{_{-\infty}}^{^{+\infty}}\mbox{$dz$} \,\left|\phi_{\beta}(z,t;\theta)\right|^{\2}
=
\mbox{$\frac{\sin^{\2}{[2\theta]}}{2}$}\left\{\, 1 - \mbox{\sc Sfo}(t) \, \right\},
\label{1}
\end{eqnarray}
where $\mbox{\sc Sfo}(t)$ represents the mass-eigenstate interference term given by
\begin{equation}
\mbox{\sc Sfo}(t) = Re
 \left[\, \int_{_{-\infty}}^{^{+\infty}}dz
\,\phi^{\dagger}_{\1}(z,t) \, \phi_{\2}(z,t) \, \right]\,
.
\label{2}
\end{equation}
Let us consider mass-eigenstate wave packets given at time $t = 0$ by
\begin{equation}
\phi_{\i}(z,0) = \left(\frac{2}{\pi a^{\2}}\right)^{ \frac{1}{4}} \exp{\left[- \frac{z^{\2}}{a^{\2}}\right]} \exp{[i p_{\i} \, z]},
\label{3}
\end{equation}
where $s = 1,\, 2$.
The wave functions which describe their time evolution are
\begin{eqnarray}
\phi_{\i}(z,t) =
\int_{_{-\infty}}^{^{+\infty}}\frac{dp_{\z}}{2 \pi} \,
\varphi(p_{\z} - p_{\i}) \exp{\left[-i\,E(p_{\z}, m_{\i})\,t +i \, p_{\z}
\,z\right]},
\label{4}
\end{eqnarray}
where
\begin{equation}
E(p_{\z}, m_{\i}) = \left(p_{\z}^{\2} + m_{\i}^{\2}\right)^{ \frac{1}{2}}
~~~~\mbox{and}~~~~
\varphi(p_{\z} - p_{\i}) =  \left(2 \pi a^{\2} \right)^{ \frac{1}{4}} \exp{\left[- \frac{(p_{\z} - p_{\i})^{\2}\,a^{\2}}{4}\right]}.
\nonumber
\end{equation}
In order to obtain the oscillation probability, we can calculate the interference term $\mbox{\sc Sfo}(t)$
by solving the following integral
\begin{eqnarray}
&&
\int_{_{-\infty}}^{^{+\infty}}\frac{dp_{\z}}{2 \pi}
\, \varphi(p_{\z} - p_{ 1}) \varphi(p_{\z} - p_{ 2})
\exp{[-i \, \Delta E(p_{\z}) \, t]}
= \nonumber\\
&&
~~~~~~~~~~~~~~~~~~~~~~~~ \exp{\left[- \frac{(a \, \Delta{p})^{\2}}{8}\right]}
\,\int_{_{-\infty}}^{^{+\infty}}\frac{dp_{\z}}{2 \pi}
\, \varphi^{\2}(p_{\z} - p_{\0})\exp{[-i \, \Delta E(p_{\z}) \, t]}, \label{6}
\end{eqnarray}
where we have changed the $z$-integration into a $p_{\z}$-integration
and introduced the quantities
\[\Delta p = p_{ \1} - p_{ \2}
\, , \, \, \, \,\,\,\,
p_{\0} = \frac{1}{2}(p_{ \1} + p_{ \2})
~~~\mbox{and}~~~
\Delta E(p_{\z}) = E(p_{\z}, m_{\1}) - E(p_{\z}, m_{\2})
. \]
The oscillation term is
bounded by the exponential function of $a \, \Delta p$ at
any instant of time. Under this condition we could never observe a
{\em pure} flavor-eigenstate. Besides, oscillations are
considerably suppressed if $a \, \Delta p > 1$. A necessary
condition to observe oscillations is that $a \, \Delta p \ll 1$.
This constraint can also be expressed by $\delta p \gg \Delta p$
where $\delta p$ is the momentum uncertainty of the particle. The
overlap between the momentum distributions is indeed relevant only
for $\delta p \gg \Delta p$. Consequently, without loss of
generality, we can assume
\begin{equation}
\mbox{\sc Sfo}(t) = Re \left\{\int_{_{-\infty}}^{^{+\infty}}\frac{dp_{\z}}{2
\pi}
 \, \varphi^{\2}(p_{\z} - p_{\0})\exp{[-i \, \Delta E(p_{\z}) \, t]} \, \right\}
\label{9}.
\end{equation}

In litterature, this equation is often obtained by assuming two
mass eigenstate wave packets described by the ``same'' momentum
distribution centered around the average momentum
$\bar{p} = p_{\0}$. This simplifying hypothesis  also guarantees
{\em instantaneous} creation of a {\em pure} flavor
eigenstate {\boldmath$\nu_\alpha$} at $t = 0$ \cite{DeL04}. In fact, for $
\phi_{\1}(z,0)=\phi_{\2}(z,0)$ we get from Eq.~(\ref{0})
\begin{equation}
\phi_{\alpha}(z,0,\theta) = \left(\frac{2}{\pi
a^{\2}}\right)^{\frac{1}{4}} \exp{\left[- \frac{z^{\2}}{a^{\2}}\right]}
\exp{[i  p_{\0} \,z]}
~~~\mbox{and}~~~
\phi_{\beta}(z,0,\theta) =0 \label{9B}.
\end{equation}

In order to obtain an expression for $\phi_{\i}(z,t)$ by analytically solving the integral in Eq.~(\ref{4}) we firstly rewrite the energy $E( p_{\z}, m_{\i})$ as
\begin{eqnarray}
E( p_{\z}, m_{\i}) & = & E_{\i} \left[1 + \frac{ p_{\z}^{\2} - p_{\0}^{\2}}{E_{\i}^{\2}}\right]^{ \frac{1}{2}}
            = E_{\i} \left[1 + \sigma_{\i} \left(\sigma_{\i} + 2 \mbox{v}_{\i}\right)\right]^{ \frac{1}{2}},
\label{11}
\end{eqnarray}
where \[ E_{\i} = (m_{\i}^{\2} + p_{\0}^{\2})^{ \frac{1}{2}}\, ,\, \, \, \,
\mbox{v}_{\i} = \frac{ p_{\0}}{E_{\i}}\, \, \, \, \mbox{and}\, \, \, \,
\sigma_{\i} = \frac{ p_{\z} -  p_{\0}}{E_{\i}}\,.\]
The use of free {\em gaussian} wave
packets is justified in non-relativistic quantum mechanics because
the calculations can be carried out
exactly for these particular functions.
The reason lies in the fact that the frequency components
of the mass-eigenstate wave packets,
$E( p_{\z},m_{\i})= p_{\z}^{\2}/2 m_{\i}$, modify the momentum
distribution into ``generalized'' {\em gaussian}, easily integrated by
well known methods of analysis. The term $ p_{\z}^{\2}$ in
$E( p_{\z},m_{\i})$ is then responsible for the variation in time
of the width of the mass-eigenstate wave packets, the so-called
{\em spreading} phenomenon. In relativistic quantum mechanics the
frequency components of the mass-eigenstate wave packets,
$E( p_{\z},m_{\i})=\sqrt{ p_{\z}^{\, \2} + m_{\i}^{\2} }$, do not
permit  an immediate analytic integration. This difficulty,
however, may be remedied by assuming a sharply peaked
momentum distribution, i. e. $(a \, E_{\i})^{-1}\sim\sigma_{\i} \ll 1$.
Meanwhile, the integral in Eq.~(\ref{4}) can be {\em
analytically} solved only if we consider terms up to order
$\sigma_{\i}^2$ in the series expansion.
In this case, we can conveniently
truncate the power series
\begin{eqnarray}
E( p_{\z}, m_{\i}) & = & E_{\i} \left[1 + \sigma_{\i} \mbox{v}_{\i}  + \frac{\sigma_{\i}^{\2}}{2}\left(1 - \mbox{v}_{\i}^{\2} \right)\right] + \mathcal{O}(\sigma_{\i}^{\3})
 \approx 
  E_{\i} +  p_{\0} \sigma_{\i} + \frac{m_{\i}^{\2}}{2E_{\i}} \sigma_{\i}^{\2}.
\label{12}
\end{eqnarray}
and get an analytic expression for the
oscillation probability.
The zero-order term in the previous expansion, $E_{\i}$, gives the
standard plane wave oscillation phase. The first-order term, $ p_{\0}
\sigma_{\i}$,   will be responsible for the {\em slippage}
 due to the different group velocities of the
mass-eigenstate wave packets and represents a linear correction to
the standard oscillation phase \cite{DeL04}. Finally, the
second-order term, $\frac{m_{\i}^2}{2E_{\i}} \sigma_{\i}^2$, which is a
(quadratic) secondary correction will give the well-known
{\em spreading} effects in the time propagation of the wave packet and
will be also responsible for a {\em new} additional phase to be
computed in the final calculation. In the case of {\em gaussian}
momentum distributions for the mass-eigenstate wave packets, these
terms can all be {\em analytically} quantified \cite{Ber04}.
By substituting (\ref{12}) in Eq.~(\ref{4}) and changing the
$ p_{\z}$-integration into a $\sigma_{\i}$-integration, we obtain the
explicit form of the mass-eigenstate wave packet time evolution,
\begin{eqnarray}
\phi_{\i}(z,t)  & \approx &(2 \pi\, a^{\2})^{ \frac{1}{4}}\exp{[-i(E_{\i} \, t -  p_{\0} \, z)]}
\nonumber\\ &&~~~~~~~~~~~~~~\times~
\int_{_{-\infty}}^{^{+\infty}}\frac{d\sigma_{\i}}{2\pi}
\, E_{\i}\,
\exp{\left[-\frac{a^{\2}\, E_{\i}^{\2} \,\sigma_{\i}^{\2}}{4}\right]} \exp{\left[-i\,( p_{\0}\,t - E_{\i} \,z)\sigma_{\i} -i\,\frac{m_{\i}^{\2} \, t}{2 E_{\i}}\sigma_{\i}^{\2}\right]}\nonumber\\
                      & = &\left[\frac{2}{\pi \,a^{\2}_{\i}(t)}\right]^{ \frac{1}{4}}\exp{[-i\,(E_{\i} \, t -  p_{\0} \, z)]}\,
\exp{\left[-\frac{(z - \mbox{v}_{\i} \,t)^{\2}}{a_{\i}^{\2}(t)} -i \,\theta_{\i}(t, z)\right]},
\label{13}
\end{eqnarray}
where
\[a_{\i}(t) = a \left(1 + \frac{4\, m_{\i}^{\4}}{a^{\4}\, E_{\i}^{\6}}\,t^{\2}\right)^{ \frac{1}{2}}
~~~~\mbox{and}~~~~
\theta_{\i}(t, z) = \left\{\frac{1}{2}\arctan{\left[\frac{2\,m_{\i}^{\2}\, t}{a^{\2}\, E_{\i}^{\3}}\right]} - \frac{2\, m_{\i}^{\2}\, t} {a^{\2}\, E_{\i}^{\3}}\,\frac{(z - \mbox{v}_{\i} \,t)^{\2}}{a_{\i}^{\2}(t)}\right\}.\]
The time-dependent quantities $a_{\i}(t)$ and $\theta_{\i}(t, z)$ contain all the physically significant information \cite{Ber04} which arise from the second order term in the power series expansion (\ref{12}).
By solving the integral (\ref{9}) with the approximation (\ref{11})
and performing some mathematical manipulations, we obtain
\begin{equation}
\mbox{\sc Sfo}(t) = \mbox{\sc Bnd}(t) \times \mbox{\sc Osc}(t),
\label{20}
\end{equation}
where we have factored the time-vanishing bound of the interference term given by
\begin{equation}
\mbox{\sc Bnd}(t) = \left[1 + \mbox{\sc Sp}^{\2}(t) \right]^{-\frac{1}{4}}
\exp{\left[-\frac{(\Delta \mbox{v} \, t)^{\2}}{2a^{\2}\left[1 + \mbox{\sc Sp}^{\2}(t)\right]}\right]}
\label{21}
\end{equation}
and the time-oscillating character of the flavor conversion formula given by
\begin{eqnarray}
\mbox{\sc Osc}(t) &=& Re \left\{\exp{\left[-i\Delta E \, t -i \Theta(t)\right]} \right\}
= \cos{\left[\Delta E \, t + \Theta(t)\right]},
\label{22A}
\end{eqnarray}
where
\begin{equation}
\mbox{\sc Sp}(t) = \frac{t}{a^{\2}}\Delta\left(\frac{m^{\2}}{E^{\3}}\right) = \rho\, \frac{\Delta \mbox{v}\, t}{a^{\2} \,  p_{\0}}
~~~~\mbox{and}~~~~
\Theta(t) = \left[\frac{1}{2}\arctan{\left[\mbox{\sc Sp}(t)\right]} - \frac{a^{\2} \,  p_{\0}^{\2}}{2 \rho^{\2}}\frac{\mbox{\sc Sp}^{\3}(t)}{\left[1 + \mbox{\sc Sp}^{\2}(t)\right]}\right],
\label{24A}
\end{equation}
with
\begin{equation}
\rho = 
 1 - \left[3 + \left(\frac{\Delta E}{\bar{E}}\right)^{\2}\right] \frac{ p_{\0}^{\2}}{\bar{E}^{\2}}
 ~~~~~~ \mbox{and} ~~~~~~\bar{E} = \sqrt{E_{\1} \, E_{\2}}.
\label{241}
\end{equation}
The time-dependent quantities $\mbox{\sc Sp}(t)$ and $\Theta(t)$
carry the second order corrections and, consequently, the
{\em spreading} effect to the oscillation probability formula.
If $\Delta E \ll \bar{E}$, the parameter $\rho$ is limited by
the interval $[1,-2]$ and it assumes the zero value when $\frac{ p_{\0}^{\2}}{\bar{E}^{\2}} \approx \frac{1}{3}$.
Therefore, by considering increasing values of $ p_{\0}$,
from non-relativistic (NR) to ultra-relativistic (UR) propagation regimes,
and fixing $\frac{\Delta E}{a^{\2} \, \bar{E}^{\2}}$,
the time derivatives of $\mbox{\sc Sp}(t)$ and $\Theta(t)$ have their
signals inverted when $\frac{ p_{\0}^{\2}}{\bar{E}^{\2}}$ reaches the value $\frac{1}{3}$.
The {\em slippage} between the mass-eigenstate wave packets is
quantified by the vanishing behavior of $\mbox{\sc Bnd}(t)$.
In order to compare $\mbox{\sc Bnd}(t)$ with the correspondent
function without the second order corrections (without {\em spreading}),
\begin{equation}
\mbox{\sc Bnd}_{\mbox{\tiny $WS$}}(t) = \exp{\left[-\frac{(\Delta \mbox{v} \, t)^{\2}}{2a^{\2}}\right]},
\label{23A}
\end{equation}
we substitute ${\mbox{\sc Sp}(t)}$ given by the expression (\ref{22A})
in Eq.~(\ref{21}) and we obtain the ratio
\begin{eqnarray}
\frac{\mbox{\sc Bnd}(t)}{\mbox{\sc Bnd}_{\mbox{\tiny $WS$}}(t)}
&=& \mbox{$\left[1 + \rho^{\2} \left(\frac{\Delta E \, t}{a^{\2} \, \bar{E}^{\2}}\right)^{\2} \right]^{-\frac{1}{4}}$}
\mbox{$ \exp{\left[\frac{\rho^{\2} \,  p_{\0}^{\2} \,
\left(\Delta E \, t\right)^{\4}}
{2\,a^{\6} \, \bar{E}^{\8}\left[1 + \rho^{\2} \left(\frac{\Delta E \, t}{a^{\2} \, \bar{E}^{\2}}\right)^{\2}\right]}
\right]}.$}
\label{23AA}
\end{eqnarray}
The NR limit is obtained by setting $\rho^{\2} = 1$ and $ p_{\0} = 0$ in Eq.~(\ref{23A}).
In the same way, the UR limit is obtained by setting $\rho^{\2} = 4$
and $ p_{\0} = \bar{E}$.
In fact, the minimal influence due to second order corrections occurs
when $\frac{ p_{\0}^{\2}}{\bar{E}^{\2}} \approx \frac{1}{3}$ ($\rho \approx 0$).
Returning to the exponential term of Eq.~(\ref{21}), we observe that the oscillation amplitude is more
relevant when $\Delta \mbox{v} \, t \ll a$.
It characterizes the {\em minimal slippage} between the mass-eigenstate
wave packets which occur when the
complete spatial intersection between themselves starts to diminish
during the time evolution.
Anyway, under {\em minimal slippage} conditions, we always have
$\frac{\mbox{\sc Bnd}(t)}{ \mbox{\sc Bnd}_{\mbox{\tiny $WS$}}(t)} \approx 1$.

\begin{figure}[h]
\begin{center}
\epsfig{file= 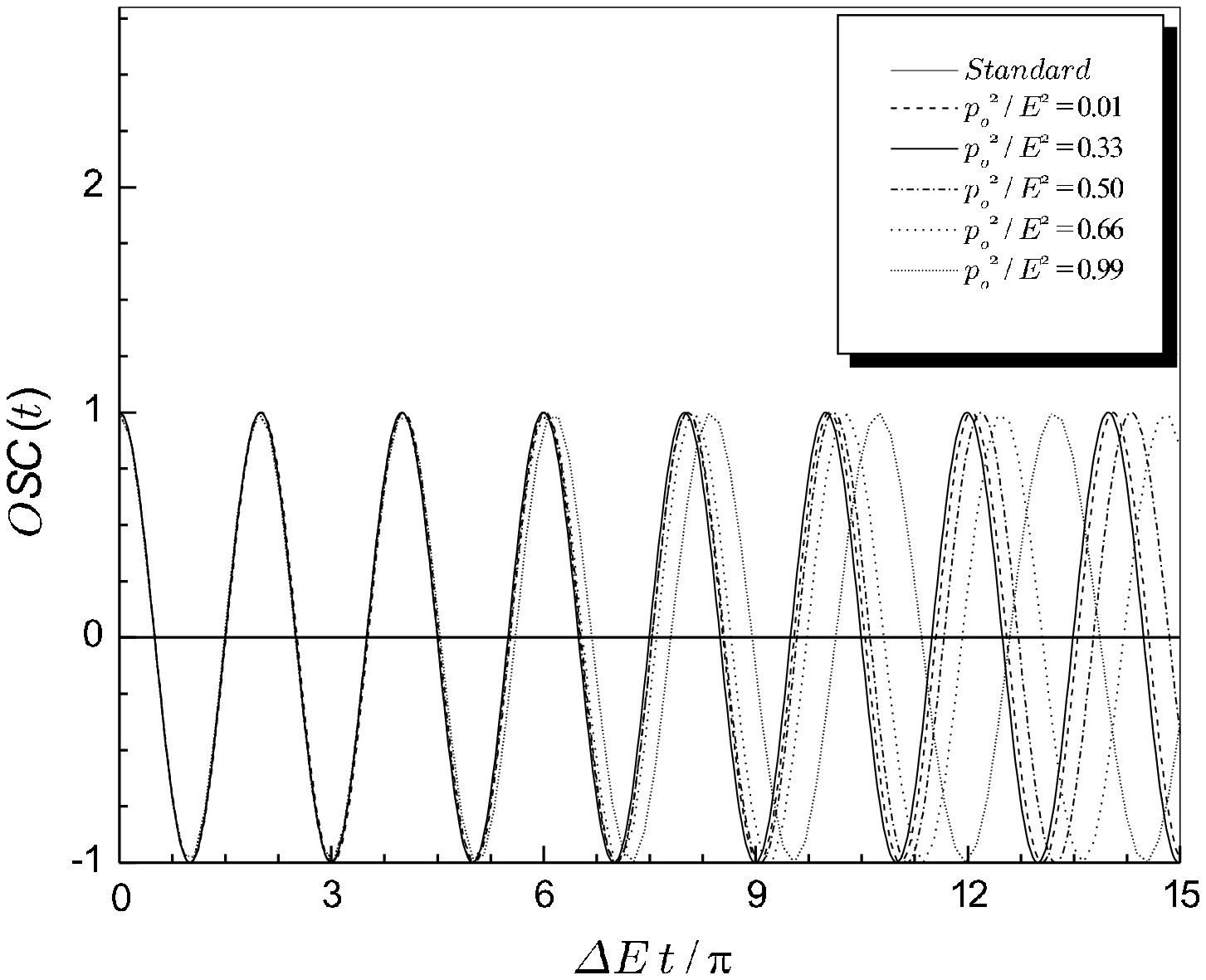, height= 11 cm, width= 14 cm}
\end{center}
\caption{\label{an4} The time-behavior of $\mbox{\sc Osc}(t)$ compared with the {\em standard} plane wave oscillation given by $\cos{[\Delta E \, t]}$
for different propagation regimes.
The additional phase $\Theta(t)$ changes the oscillating character after some time of propagation.
The maximal deviation occurs for $\frac{p_o^{\2}}{\bar{E}^{\2}} \approx \frac{1}{3}$.}
\end{figure}

The oscillating function $\mbox{\sc Osc}(t)$ of the interference
term $\mbox{\sc Sfo}(t)$ differs from the {\em standard} oscillating
term, $ \cos{[\Delta E \, t]}$,
by the presence of the additional phase $\Theta(t)$
which is essentially a second order correction.
The modifications introduced by the additional phase $\Theta(t)$ are discussed in Fig.~\ref{an4} \cite{Ber04}
where we have compared the time-behavior of $\mbox{\sc Osc}(t)$ to $\cos{[\Delta E \, t]}$ for different propagation regimes.
The bound {\em  effective} value assumed by $\Theta (t)$
is determined by the vanishing behavior of $\mbox{\sc Bnd}(t)$.

\begin{figure}[h]
\begin{center}
\epsfig{file= 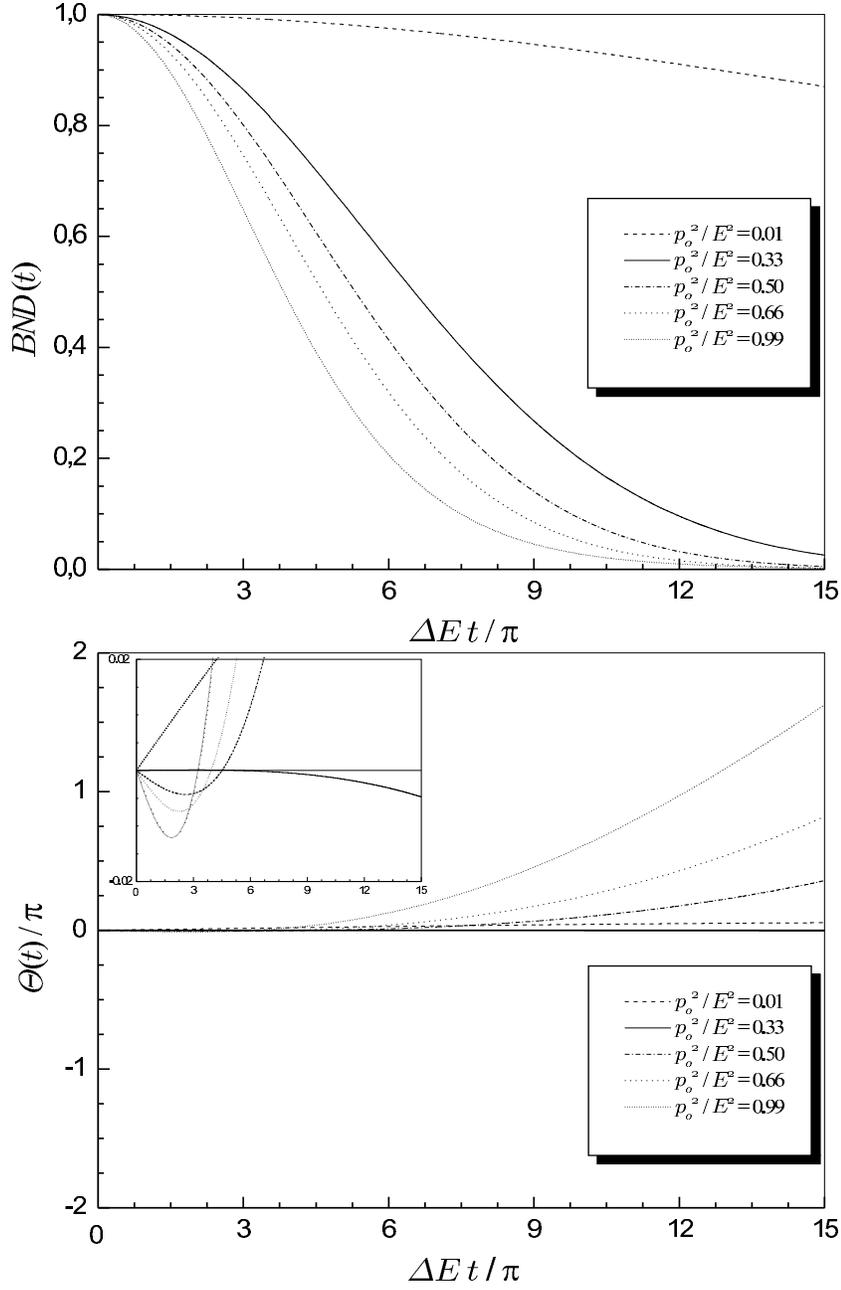, height= 17 cm, width= 11 cm}
\end{center}
\caption{\label{an5} The time-behavior of the additional phase $\Theta(t)$.
The values assumed by $\Theta (t)$ are {\em  effective} while the interference term does not vanish.
In the upper box we can observe the behavior of $\mbox{\sc Bnd}(t)$ which determines the limit values effectively assumed by
$\Theta(t)$ for each propagation regime.
For relativistic regimes with $\frac{ p_{\0}^{\2}}{\bar{E}^{\2}} > \frac{1}{3}$, the function $\Theta(t)$ rapidly reaches its lower limit as we can observe in the small box above.
We have used $a \, \bar{E} = 10$.}
\end{figure}

To illustrate this scalar flavor oscillation behavior, we plot both the curves representing $\mbox{\sc Bnd}(t)$ and $\Theta(t)$ in Fig.~\ref{an5} \cite{Ber04}.
We note the phase slowly changing in the NR regime.
The modulus of the phase $|\Theta(t)|$ rapidly reaches its upper limit when $\frac{ p_{\0}^{\2}}{\bar{E}^{\2}} > \frac{1}{3}$ and, after a certain time, it continues to evolve approximately linearly in time.
But, effectively, the oscillations rapidly vanishes.
By superposing the effects of $\mbox{\sc Bnd}(t)$ in Fig.~\ref{an5} and
the oscillating character $\mbox{\sc Osc}(t)$ expressed in Fig.~\ref{an4}, we immediately obtain the flavor oscillation probability which is explicitly given by
\begin{eqnarray}
P_{\scalar}(\mbox{\boldmath$\nu_\alpha$}\rightarrow\mbox{\boldmath$\nu_\beta$};t)
 \approx
\mbox{$\frac{\sin^{\2}{[2\theta]}}{2}$}\left\{1 - \left[1 + \mbox{\sc Sp}^{\2}(t) \right]^{-\frac{1}{4}}\right.
\left.\exp{\left[-\frac{(\Delta \mbox{v} \, t)^{\2}}{2a^{\2}\left[1 + \mbox{\sc Sp}^{\2}(t)\right]}\right]}
\cos{\left[\Delta E \, t + \Theta(t)\right]}
  \right\}.&
\label{25A}
\end{eqnarray}
Obviously, the larger is the value of $a \, \bar{E}$, the smaller are the wave packet effects.
If it was sufficiently larger to not consider the second order corrections expressed in Eq.~(\ref{11}),
we could  compute the oscillation probability with the leading corrections due to the {\em slippage} effect,
\begin{eqnarray}
P_{\scalar}(\mbox{\boldmath$\nu_\alpha$}\rightarrow\mbox{\boldmath$\nu_\beta$};t) \approx
 \mbox{$\frac{\sin^{\2}{[2\theta]}}{2}$}
 \left\{1- \exp{\left[-\frac{(\Delta \mbox{v} \, t)^{\2}}{2\, a^{\2}}\right]}\cos{[\Delta E \, t ]}\right\}&
\label{20AA}
\end{eqnarray}
which corresponds to the same result obtained by \cite{DeL04}.
By assuming an UR propagation regime with $t \approx L$ and $E_i \sim p_{\0}$,
under {\em minimal slippage} conditions ($\Delta \mbox{v} \, L \ll a$), the above expression reproduces the {\em standard} plane wave result,
\begin{eqnarray}
P_{\scalar}(\mbox{\boldmath$\nu_\alpha$}\rightarrow\mbox{\boldmath$\nu_\beta$};L)
&\approx& \mbox{$\frac{\sin^{\2}{[2\theta]}}{2} \left\{1- \left[1 -\frac{(\Delta \mbox{v} \, L)^{\2}}{2\, a^{\2}}\right] \cos{\left[\Delta E \, L \right]}\right\}$}\nonumber\\
&\approx& \mbox{$\frac{\sin^{\2}{[2\theta]}}{2}\left\{1- \cos{\left[\frac{\Delta m^{\2}}{4 p_{\0}} \, L \right]}\right\}$},\nonumber\\
&=& \mbox{$\sin^{\2} [2 \theta] \sin^{\2} \left[ \frac{\Delta m^{\2}}{4 p_{\0}} \, L \right]$}.
\label{20A}
\end{eqnarray}
since we have assumed $a \, \bar{E} \gg 1$.


\section*{III. Dirac Formalism}

The results in the previous section have been obtained by considering {\em scalar} mass-eigenstates.
Neutrinos are, however, {\em fermions}.
The time evolution of a spin one-half particle must be described by the Dirac equation.
To introduce the {\em fermionic} character in the study of quantum oscillation phenomena, we shall use the Dirac equation as the evolution equation for the mass-eigenstates.
The Eq.~(\ref{0}) now becomes
\begin{eqnarray}
\Psi(z,t) &=& \psi_{\1}(z,t)\cos{\theta}\,\mbox{\boldmath$\nu_{\1}$} + \psi_{\2}(z,t)\sin{\theta}\,\mbox{\boldmath$\nu_{\2}$}\nonumber\\
          &=& \left[\psi_{\1}(z,t)\cos^2{\theta} + \psi_{\2}(z,t)\sin^2{\theta}\right]\,\mbox{\boldmath$\nu_\alpha$} + \left[\psi_{\1}(z,t) - \psi_{\2}(z,t)\right]\cos{\theta}\sin{\theta}\,\mbox{\boldmath$\nu_\beta$}\nonumber\\
          &=& \psi_{\alpha}(z,t;\theta)\,\mbox{\boldmath$\nu_\alpha$} + \psi_{\beta}(z,t;\theta)\,\mbox{\boldmath$\nu_\beta$},
\label{0B}
\end{eqnarray}
where $\psi_i(z,t)$ satisfies the Dirac equation for a mass $m_i$.
The natural extension of Eq.~(\ref{9B}) reads
\begin{equation}
\psi_{\alpha}(z,0,\theta) = \phi_{\alpha}(z,0,\theta) \, w
\label{22}
\end{equation}
where $w$ is a constant spinor which satisfies the normalization condition $w^{\dagger} w = 1$.

\subsection*{III.1.  Dirac wave packets and the oscillation formula}

To describe the time evolution of mass-eigenstate Dirac wave packets,
we could be inclined to superpose only positive frequency solutions of the Dirac equation.
It seems, at first glance, a reasonable choice.
However, when the initial state has the form given in Eq.~(\ref{22}), it is necessary to
superpose both positive and negative frequency solutions of Dirac equation.
In order to clear up this point,
let us express the flavor state $\psi_{\alpha}(z,t,\theta)$ in terms of
\begin{eqnarray}
\psi_i(z,t)
&=& \int_{_{-\infty}}^{^{+\infty}}\frac{d p_{\z}}{2\pi} \exp{[ i  p_{\z} z]}\sum_{i=1,2}\{b^s( p_{\z},m_{\i})\,u^s( p_{\z},m_{\i}) \exp{[-iE( p_{\z},m_{\i}) t]}\nonumber\\
&&~~~~~~~~~~~~~~~~~~~~~~~~~~~~~~~~~~~~~+ d^{s*}(\mi  p_{\z},m_{\i})\,v^s(\mi  p_{\z},m_{\i}) \exp{[+iE( p_{\z},m_{\i})t]}\}.
\label{23}
\end{eqnarray}
At time $t=0$ the mass-eigenstate wave functions satisfy $\psi_{\1}(z,0)=\psi_{\2}(z,0)$ (this
guarantees the {\em instantaneous} creation of a {\em pure} flavor-eigenstate {\boldmath$\nu_\alpha$} as we have appointed in section II).
The Fourier transform of $\psi_i(z,0)$ is
\begin{eqnarray}
\sum_{i=1,2}\left[b^{s}( p_{\z},m_{\i})\,u^{s}( p_{\z},m_{\i}) + d^{s*}(\mi  p_{\z},m_{\i})\,v^{s}(\mi  p_{\z},m_{\i})\right].
\label{24}
\end{eqnarray}
By observing that the Fourier transform of $\phi_{\alpha}(z,0,\theta)$ is given by $\varphi( p_{\z} -  p_{\0})$ (see Eq.~(\ref{9B})), we immediately obtain the Fourier transform of $\psi_{\alpha}(z,0,\theta)$,
\begin{eqnarray}
\varphi( p_{\z} -  p_{\0}) w & = &\sum_{i=1,2}\left[b^{s}( p_{\z},m_{\i})\,u^{s}( p_{\z},m_{\i}) + d^{s*}(\mi  p_{\z},m_{\i})\,v^{s}(\mi  p_{\z},m_{\i})\right].
\label{25}
\end{eqnarray}
Using the orthogonality properties of Dirac spinors, we find \cite{Zub80}
\begin{eqnarray}
b^s( p_{\z},m_{\i}) &=& \varphi( p_{\z} -  p_{\0})\,u^{s \dagger}( p_{\z},m_{\i}) \, w, \nonumber\\
d^{s*}(\mi  p_{\z},m_{\i}) &=& \varphi( p_{\z} -  p_{\0})\,v^{s \dagger}(\mi  p_{\z},m_{\i}) \, w.
\label{26}
\end{eqnarray}
These coefficients carry an important physical information.
For {\em any} initial state which has the form given in Eq.~(\ref{22}), the negative frequency solution coefficient $d^{s*}(\mi  p_{\z},m_{\i})$ necessarily provides a non-null contribution to the time evolving wave packet.
This obliges us to take the complete set of Dirac equation solutions to construct the wave packet.
Only if we consider a momentum distribution given by a delta function (plane wave limit) and suppose an initial spinor $w$ being a positive energy mass-eigenstate with momentum $ p_{\0}$, the contribution due to the negative frequency solutions $d^{s*}(\mi  p_{\z},m_{\i})$ will be null.

Having introduced the Dirac wave packet prescription, we are now in a position to calculate the flavor conversion formula.
The following calculations do not depend on the gamma matrix representation.
By substituting the coefficients given by Eq.~(\ref{26}) in Eq.~(\ref{23}) and using the well-known spinor properties \cite{Zub80},
\begin{eqnarray}
&&\sum_{i=1,2}u^s( p_{\z},m_{\i})\overline{u}^s( p_{\z},m_{\i}) = \frac{\gamma^0 E( p_{\z},m_{\i}) - \gamma^3  p_{\z} + m_i}{2E( p_{\z},m_{\i})}, \nonumber\\
&&\sum_{i=1,2}v^s(\mi  p_{\z},m_{\i})\overline{v}^s(\mi  p_{\z},m_{\i}) = \frac{\gamma^0 E( p_{\z},m_{\i}) + \gamma^3  p_{\z} -m_i}{2E( p_{\z},m_{\i})},
\label{28}
\end{eqnarray}
we obtain
\begin{eqnarray}
\psi_i(z,t) & = &\int_{_{-\infty}}^{^{+\infty}}\frac{d p_{\z}}{2 \pi} \, \varphi( p_{\z} - p_{\0}) \, \exp{[i p_{\z} z]} \left\{\cos{[E( p_{\z},m_{\i}) t]} -\frac{i\gamma^0\left(\gamma^3  p_{\z}+ m_i\right)}{E( p_{\z},m_{\i})}\sin{[E( p_{\z},m_{\i}) t]}\right\} \, w.~~~~~
\label{29}
\end{eqnarray}
By simple mathematical manipulations, the new interference oscillating term will be written as
\begin{eqnarray}
\mbox{\sc Dfo}(t) & = &\int_{_{-\infty}}^{^{+\infty}}\frac{d p_{\z}}{2 \pi} \, \varphi^2( p_{\z} - p_{\0})
\left\{ \, \left[ \, 1 -
f( p_{\z},m_{\1},m_{\2}) \, \right] \, \cos [
\epsilon_{\mi}( p_{\z},m_{\1},m_{\2}) \, t ] \right. + \nonumber
\\  & & ~~~~~~~~~~~~~~~~~~~~~~~~~~~~~~~~~~~~~~~~~~~~~~~~ \left. f( p_{\z},m_{\1},m_{\2}) \, \cos [
\epsilon_{\pl}( p_{\z},m_{\1},m_{\2}) \, t] \right\}
\label{30}
\end{eqnarray}
where
\[
f( p_{\z},m_{\1},m_{\2})= \frac{E( p_{\z},m_{\1})E( p_{\z},m_{\2}) -
 p_{\z}^{\2} - m_{\1} m_{\2}}{2\, E( p_{\z},m_{\1})E( p_{\z},m_{\2})}
~~~\mbox{and}~~~
\epsilon_{\pm}( p_{\z},m_{\1},m_{\2})=E( p_{\z},m_{\1}) \pm
E( p_{\z},m_{\2})~.
 \]
The time-independent term $f ( p_{\z},m_{\1},m_{\2})$ deserves
some comments. It has a minimum  at $p_{\z}=0$ and two maxima at
$p_z = \pm\sqrt{m_{\1} m_{\2}}$. It goes rapidly to zero when
$p_{\z} \gg m_{1,2}$ (ultra-relativistic limit) as well as when
$p_{\0} \ll m_{1,2}$ (non-relativistic limit). It means that when
we consider a momentum distribution sharply peaked around $p_{\0}
\gg m_{1,2}$ or $p_{\0} \ll m_{1,2}$ the corrections introduced by
$f ( p_{\z},m_{\1},m_{\2})$ are negligible. The maximum value of
$f ( p_{\z},m_{\1},m_{\2})$ is
\begin{equation}
f_{max} ( p_{\z},m_{\1},m_{\2}) = \frac{1}{2} - \frac{\sqrt{m_{\1}m_{\2}}}{m_{\1} +m_{\2}}
\label{36}
\end{equation}
which vanishes in the limit $m_{\1} = m_{\2}$. The effects
introduced by $f ( p_{\z},m_{\1},m_{\2})$ are relevant only when
$\Delta m \approx m_{\1} \gg m_{\2}$. Meanwhile, what is
interesting about the result in Eq.~(\ref{30}) is that it was
obtained without any assumption on the initial spinor $w$.
Otherwise, the initial spinor carries some fundamental physical
information about the created state. And this could be relevant in
the study of chiral oscillations \cite{DeL98} where the initial
state plays a fundamental role. In comparison with the standard
treatment of neutrino oscillations done by using {\em scalar} wave
packets, where the interference term  $\mbox{\sc Sfo}(t)$ is given
by Eq.~(\ref{9}) with $\Delta E( p_{\z}) \equiv \epsilon_{\mi}(
p_{\z},m_{\1},m_{\2})$, we note in $\mbox{\sc Dfo}(t)$ two
additional terms. In the first one, the {\em standard} oscillating
term $\cos{[\epsilon_{\mi}( p_{\z},m_{\1},m_{\2})\, t]}$, which
arises from the interference between mass-eigenstate components of
equal sign frequencies, is multiplied by a {\em new factor}
obtained by the products $u^{\dagger}( p_{\z},m_{\1})\,u(
p_{\z},m_{\2})$, $v^{\dagger}(\mi p_{\z},m_{\1})\,v(\mi
p_{\z},m_{\2})$ and h.c.. The second one is a {\em new oscillating
term}, $\cos{[\epsilon_{\pl}( p_{\z},m_{\1},m_{\2})\, t]}$, which
comes from the interference between mass-eigenstate components of
positive and negative frequencies. The factor multiplying such an
additional oscillating term is obtained by the products
$u^{\dagger}( p_{\z},m_{\1})\,v(\mi  p_{\z}, m_{\2})$,
$v^{\dagger}(\mi p_{\z},m_{\1})\,u( p_{\z},m_{\2})$ and h.c.. The
new oscillations have very high frequencies. Such a peculiar
oscillating behavior is similar to the phenomenon referred to as
{\em Zitterbewegung}. In atomic physics, the electron exhibits
this violent quantum fluctuation in the position and becomes
sensitive to an effective potential which explains the Darwin term
in the hydrogen atom \cite{Sak87}. We shall see later that, at the
instant of creation, such rapid oscillations introduce a small
modification in the oscillation formula.

Returning to the starting point, if we had postulated a wave packet
made up exclusively of positive frequency plane-wave solutions,
the oscillation term $\cos{[\epsilon_{\pl}( p_{\z},m_{\1},m_{\2})\, t]}$ would have vanished.
It reinforces the argument that, in constructing  Dirac wave
packets, we cannot simply forget the contributions of negative
frequency components.

\subsection*{III.2. First order modifications to the oscillation formula}

A more satisfactory interpretation of the modifications introduced by the Dirac formalism is given when we
explicitly calculate $\mbox{\sc Dfo}(t)$.
By considering the energy $E(p_{\z},m_{\i})$ expansion
up to the second order terms in Eq.~(\ref{12}), we
include an analysis of {\em spreading} effects.
In this preliminary
study, we are, however, interested only to first order corrections.
Thus, we  approximate the frequency components by
\begin{equation}
E( p_{\z},m_{\i}) \approx \, E_{\i} \,  + \mbox{v}_{\i} \, \left(
 p_{\z}- p_{\0}  \, \right) ~.
\label{34AA}
\end{equation}
As a consequence of this approximation, we get
\begin{eqnarray}
f( p_{\z},m_{\1},m_{\2}) & \approx & \mbox{$\frac{1}{2}$} \, \left\{
1 - \mbox{v}_{\1}\mbox{v}_{\2}\left(1 + \frac{m_{\1}m_{\2}}{p_{\0}^{\2}}\right)  + \mbox{v}_{\1}\mbox{v}_{\2} \, \left[\left(\mbox{v}_{\1}^{\2} + \mbox{v}_{\2}^{\2}\right)
\left(1 +  \frac{m_{\1}m_{\2}}{p_{\0}^{\2}}\right)
-2 \right] \, \frac{ p_{\z}- p_{\0} }{ p_{\0}} \, \right\}
\end{eqnarray}
and
\begin{eqnarray}
\epsilon_{\ppm}( p_{\z},m_{\1},m_{\2}) & \approx & E_{\1} \pm
E_{\2} + \left( \, \mbox{v}_{\1} \pm \mbox{v}_{\2} \, \right) \,\left(
 p_{\z}- p_{\0}  \, \right) ~.
\end{eqnarray}
For UR particles ($m_{\i}\ll  p_{\0}$), we can also
use the following expression for the central energy values
($E_{\i}$) and the group velocities ($\mbox{v}_{\i}$)  of the
mass-eigenstate wave packets,
\[ E_{\i} \approx \,  p_{\0} +
\frac{m_{\i}^{\2}}{2\, p_{\0}}~~~~~\mbox{and} ~~~~~ \mbox{v}_{\i} \approx
\, 1 -  \frac{m_{\i}^{\2}}{2\, p_{\0}^{\, \2}}~.
\]
This implies
\begin{eqnarray*}
f(p_{\z},m_{\1},m_{\2}) & \approx & \left(\frac{\Delta m}{2 \, p_{\0}}\right)^2 \, \left( \, 1 -  2 \, \,
\frac{ p_{\z}- p_{\0} }{ p_{\0}} \, \right) \, \, , \\
\epsilon_{\pl}( p_{\z},m_{\1},m_{\2}) & \approx & 2 \,  p_{\0}   \,
\left[ \, 1 \,  + \, \frac{m_{\1}^{\2}+ m_{\2}^{\2}}{4\, p_{\0}^{\,
\2} } \, + \, \frac{ p_{\z}- p_{\0} }{ p_{\0}} \, \left( \,  1 -
\frac{m_{\1}^{\2} + m_{\2}^{\2}}{4\, p_{\0}^{\,  \2} }\, \right) \,
\right]\, \, ,\\
\epsilon_{\mi}( p_{\z},m_{\1},m_{\2}) & \approx &  \frac{\Delta
m^{\2}}{2\, p_{\0}}  \, \left[ \, 1  - \, \frac{ p_{\z}- p_{\0}
}{ p_{\0}} \, \right]\, \, .
\end{eqnarray*}
where $(\Delta m)^{\2}= (m_{\1} - m_{\2})^{\2}$ is different from $\Delta m^{\2}= m_{\1}^{\2} - m_{\2}^{\2}$
which appears in the {\em standard} oscillation phase. Finally, by simple algebraic manipulations and after {\em gaussian}
integrations, we find
\begin{eqnarray}
\mbox{\sc Dfo}(t) &\approx& \mbox{$\exp{\left[-\left(\frac{\Delta m^2\, t}{2 \sqrt{2}a  p_{\0}^2}\right)^2\right]}\left\{\left[1 - \left(\frac{\Delta m}{2 p_{\0}}\right)^2\right]\cos{\left[\frac{\Delta m^2}{2  p_{\0}}t\right]} + \left(\frac{\Delta m}{2 p_{\0}}\right)^2 \frac{\Delta m^2}{a^2 p_{\0}^3}t\sin{\left[\frac{\Delta m^2}{2  p_{\0}}t\right]}\right\}$}\nonumber\\
         &+& \mbox{$\exp{\left[-\frac{t^2}{2 a^2}\left(2 - \frac{m_{\1}^2 + m_{\2}^2}{2  p_{\0}^2}\right)^2\right]}\left(\frac{\Delta m}{2 p_{\0}}\right)^2 \left\{\cos{\left[ p_{\0}t\left(2 + \frac{m_{\1}^2 + m_{\2}^2}{2  p_{\0}^2}\right)\right]}\right.$}\nonumber\\
         & &\mbox{$\left.~~~~~~~~~~~~~~~~~~~~~~~~~~~~~~~~~~~ + \frac{2 p_{\0}t}{(a  p_{\0})^2}\left(2 - \frac{m_{\1}^2 + m_{\2}^2}{2  p_{\0}^2}\right)\sin{\left[ p_{\0}t\left(2 + \frac{m_{\1}^2 + m_{\2}^2}{2  p_{\0}^2}\right)\right]}\right\}$}.
\label{A14}
\end{eqnarray}
As we have already noticed, the oscillating functions going with the second exponential function in Eq.~(\ref{A14}) arise from the interference between positive and negative frequency solutions of the Dirac equation.
It produces very high frequency oscillations which is similar to the quoted phenomenon of {\em Zitterbewegung} \cite{Sak87}.
The oscillation length which characterizes the very high frequency oscillations is given by $L^{\tiny VHF}_{0sc} \approx \frac{2 \pi}{ p_{\0}}$.
Obviously, $L^{\tiny VHF}_{0sc}$ is much smaller than the standard oscillation length given by $L^{Std}_{0sc} = \frac{4 \pi  p_{\0}}{\Delta m^2}$.
It means that the propagating particle exhibits a violent quantum fluctuation of its flavor quantum number around a flavor average value which oscillates with $L^{Std}_{0sc}$.
Meanwhile, except at times $t \sim 0$, it provides a practically null contribution to the oscillation probability.
To explain such a statement, let us suppose that an experimental measurement takes place after a time $t \approx L$ for UR particles.
The observability conditions impose that the propagation distance $L$ must be larger than the wave packet localization $a$.
Since the (second) exponential function vanishes when $L \gg a$, for measurable distances,
the effective flavor conversion formula will not contain such very high frequency oscillation terms, and can be written as
\begin{eqnarray}
 P_{\Dirac}(\mbox{\boldmath$\nu_\alpha$}\rightarrow\mbox{\boldmath$\nu_\beta$};L)
 &\approx& \mbox{$\frac{\sin^2{[2\theta]}}{2}\left\{ 1 -\exp{\left[-\left(\frac{\Delta m^2\, L}{2 \sqrt{2}a  p_{\0}^2}\right)^2\right]}\left\{\left[1 - \left(\frac{\Delta m}{2 p_{\0}}\right)^2\right]\cos{\left[\frac{\Delta m^2}{2  p_{\0}}L\right]}~~ \right.\right.$}\nonumber\\
 & & \mbox{$\left.\left. ~~~~~~~~~~~~~~~~~~~~~~~~~~~~~~~~~~~~~~~~~~~~~ + \left(\frac{\Delta m}{2 p_{\0}}\right)^2 \frac{\Delta m^2}{a^2 p_{\0}^3}L\sin{\left[\frac{\Delta m^2}{2  p_{\0}}L\right]}\right\}\right\}$}.
\label{A14BB}
\end{eqnarray}
For distances which are restrict to the interval $a \ll L \ll a \frac{2 \sqrt{2}  p_{\0}^2}{\Delta m^2}$ we observe the {\em minimal slippage} between the wave packets.
In this case, we could suddenly approximate the oscillation probability to
\begin{eqnarray}
 P_{\Dirac}(\mbox{\boldmath$\nu_\alpha$}\rightarrow\mbox{\boldmath$\nu_\beta$};L)
 &\approx& \mbox{$\frac{\sin^2{[2\theta]}}{2}\left\{
1 - \left[1-\left(\frac{\Delta m^2 L}{2 \sqrt{2}a  p_{\0}^2}\right)^2\right] \left[1 - \left(\frac{\Delta m}{2 p_{\0}}\right)^2\right]\cos{\left[\frac{\Delta m^2}{2  p_{\0}}L\right]} \right\}$}
\label{A14B}
\end{eqnarray}
however, we reemphasize that it is {\em not} valid for $T \approx L \sim 0$ when the rapid oscillations are still relevant ($L < a$).
By comparing the result of Eq.~(\ref{A14B}) with the {\em scalar} oscillation probability of Eq.~(\ref{20}),
we notice a deviation of the order $\left(\frac{\Delta m}{2 p_{\0}}\right)^2$ that appears as an additional coefficient of the cosine function.
It is not relevant in the UR limit as we have noticed after studying the function $f(p_{\z},m_{\1},m_{\2})$.

\subsection*{III.3. A brief extension to quantum field treatment}

Now we try to establish a tenuous
correspondence between our results and the quantum field theory (QFT) treatment.
It was extensively demonstrated in the literature \cite{Ric93,Giu93,Giu02B}
that the oscillating particle cannot be treated in isolation.
The oscillation process must be considered globally: the oscillating
states become {\em intermediate} states, not directly observed,
which propagate between a {\em source}
and a {\em detector}. This idea can be implemented in QFT
when the {\em intermediate}
oscillating states are represented by internal lines of
Feynman diagrams
and the interacting particles at source/detector are described
by {\em external} wave packets \cite{Giu93,Beu03}.
In this context, let us consider the weak flavor-changing processes
occurring through the {\em intermediate} propagation of a neutrino,
\begin{equation}
 p_I \rightarrow  p_F + \alpha + \nu_{\alpha}
 ~~(oscillation)~~
 \nu_{\beta} + D_I \rightarrow \beta + D_F
\label{A15}
\end{equation}
where $ p_I$ and $ p_F$ ($D_I$ and $D_F$) are respectively
the initial and final production (detection) particles.
The amplitude for the process is represented by
\begin{equation}
\mathcal{A} = \mbox{$\left\langle  p_F, D_F \left|\mathbf{T}\left(
\exp{\left[-i\,\int{dx^{\4} \, \mathcal{H}_I}\right]}\right)
- \mathbf{1}\right|  p_I, D_I \right\rangle$}
\label{A16}
\end{equation}
where $\mathcal{H}_I$ is the interaction Hamiltonian for the
intermediate particle and $\mathbf{T}$ is the time ordering
operator.
After some mathematical manipulations \cite{Beu03},
this amplitude can be represented by the integral
\begin{eqnarray}
\mathcal{A} &=& \int{\frac{dE\, d^{\3}\mathbf{p}}{(2\pi)^{\4}}\,
F(E,\mathbf{p})}\,
G(E,\mathbf{p},t_D,t_P)\,
\exp{[i\, \mathbf{p}\cdot(\mathbf{x}_D - \mathbf{x}_P)]}
\label{A17}
\end{eqnarray}
where the function $F(E,\mathbf{p})$ represents the {\em overlap}
of the incoming and outgoing wave
packets, both at the source and at the detector, and
the {\em Green} function in the momentum space,
$G(E, \mathbf{p}, t_D, t_P)$, represents the fermion propagator which carries the
information of the oscillation process.
The overlap function is independent of production
and detection times and positions
($t_P$, $t_D$, $\mathbf{x}_P$, $\mathbf{x}_D)$ and depends on the
the directions of incoming and outgoing momenta.
In certain way, the physical conditions of source and detector,
in terms of time and space intervals, are better defined in this framework
than in the {\em intermediate} wave packet framework.
Anyway, to understand the oscillation process
we must turn back to the definition of mixing in quantum mechanics.
It is similar in field theory, except
that it applies to fields, not to physical states.
This difference allows to bypass the problems
arising in the definition of flavor and mass bases \cite{Beu03}.
In one-dimensional spatial coordinates, the mixing
is illustrated by the unitary transformation
\begin{equation}
\psi_{\sigma}(z,t;\theta) = \mathcal{G}^{\mi \1}(\theta; t)\,
\psi_i(z,t)\,\mathcal{G}(\theta; t)
\label{A00}
\end{equation}
as the result of the noncoincidence of the flavor basis
($\sigma =\,\alpha, \, \beta$)
and the mass basis ($i =\, 1,\, 2$).
The Eq.~(\ref{A00}) gives the Eq.~(\ref{0B})
when the generator of mixing transformations
$\mathcal{G}(\theta; t)$ is given by
\begin{eqnarray}
\mathcal{G}(\theta; t) &=& \exp[\theta \int\,dz \, \psi_{\1}(z,t)\psi_{\2}(z,t)
-\psi_{\2}(z,t)\psi_{\1}(z,t)]
\label{A00B}.
\end{eqnarray}
By taking the one-dimensional representation of Eq.~(\ref{A17}),
the propagator $G(E, p_{\z},t_D,t_P)$
can also be written in the flavor basis as
\begin{eqnarray}
G^{\alpha\beta}(\theta; E, p_{\z},T) &=&
 \mathcal{G}^{\mi \1}(\theta; t)\,G(E, p_{\z},T)\,\mathcal{G}(\theta; t)
=
 \mathcal{G}^{\mi \1}(\theta; t)\,G(E, p_{\z},t_D,t_P)\,\mathcal{G}(\theta; t)~~
\label{A00C}
\end{eqnarray}
with $T = t_D - t_P$.

In particular, by following the Blasone and Vitiello (BV)
prescription \cite{Bla95,Bla03B},
the definition of a Fock space of weak eigenstates becomes possible
and a nonperturbative flavor oscillation amplitude can be derived.
In this case, the complete Lagrangian (density) is split in
a propagation Lagrangian,
\begin{eqnarray}
\mathcal{L}_{p}(z,t) &=&
\bar{\psi}_{\1}(z,t)\,\left(i \,\partial\hspace{-0.2cm}\slash\hspace{0.1cm} - m_{\1}\right)\,\psi_{\1}(z,t)
+\bar{\psi}_{\2}(z,t)\,\left(i \,\partial\hspace{-0.2cm}\slash\hspace{0.1cm} - m_{\2}\right)\,\psi_{\2}(z,t),
\label{A21}
\end{eqnarray}
and an interaction Lagrangian
\begin{eqnarray}
\mathcal{L}_{i}(z,t)&=&
\bar{\psi}_{\alpha}(z,t;\theta)\,\left(i \,\partial\hspace{-0.2cm}\slash\hspace{0.1cm} - m_{\alpha}\right)\,\psi_{\alpha}(z,t;\theta)
+\bar{\psi}_{\beta}(z,t;\theta)\,\left(i \,\partial\hspace{-0.2cm}\slash\hspace{0.1cm} - m_{\beta}\right)\,\psi_{\beta}(z,t;\theta)
\nonumber\\&&~~~~~~~~~~~~~~~~~~~~~~~~~~~~~~~
- m_{\alpha\beta} \,\left(\bar{\psi}_{\alpha}(z,t;\theta)\psi_{\beta}(z,t;\theta)
+ \bar{\psi}_{\beta}(z,t;\theta)\psi_{\alpha}(z,t;\theta)\right),
\label{A20}
\end{eqnarray}
where
\begin{eqnarray}
m_{\alpha (\beta)} = m_{\1(\2)}\, \cos^{\2}{\theta} + m_{\2(\1)}\,\sin^{\2}{\theta}
~~~\mbox{and}~~~
m_{\alpha\beta} = (m_{\1} - m_{\2})\, \cos{\theta}\sin{\theta}.\nonumber
\end{eqnarray}
In general, the two subsets of the Lagrangian can be distinguished if there is a flavor
transformation which is a symmetry of $\mathcal{L}_{i}(z,t)$ but not of $\mathcal{L}_{p}(z,t)$.
Particle mixing occurs if the propagator built from $\mathcal{L}_{p}(z,t)$,
and representing the creation of a particle of flavor $\alpha$ at point
$z$ and the annihilation of a particle of flavor $\beta$ at point $z^{\prime}$,
is not diagonal, i.e. not zero for $\beta = \alpha$.
The free fields $\psi_i(z,t)$ can be quantized in the usual way by rewriting
the momentum distributions $b^s( p_{\z},m_{\i})$ and $d^{s*}(\mi  p_{\z},m_{\i})$ in Eq.~(\ref{23})
as creation and annihilation operators
${\sc B}^s( p_{\z},m_{\i})$ and ${\sc D}^{s\dagger}(\mi  p_{\z},m_{\i})$.
The interacting fields are then given by
\begin{eqnarray}
\psi_{\sigma}(z,t)&=& \mbox{$\int_{_{\infm}}^{^{\infp}}\frac{d p_{\z}}{2\pi} \exp{[i  p_{\z} z]}$}
\sum_{i=1,2}\{{\sc B}^s_{\sigma}( p_{\z}; t)\,u^s_{\sigma}( p_{\z}; t)
        + {\sc D}^{s*}_{\sigma}(\mi  p_{\z}; t)\,v^s_{\sigma}(\mi  p_{\z}; t)\}
\label{A22}
\end{eqnarray}
where the new flavor creation and annihilation operators which satisfy canonical
anticommutation relations are defined by means of Bogoliubov
transformations \cite{Bla03B} as
\begin{equation}
{\sc B}^s_{\sigma}( p_{\z}; t) = \mathcal{G}^{\mi \1}(\theta; t)\,{\sc B}^s( p_{\z},m_{\i})\,\mathcal{G}(\theta; t)
~~~\mbox{and}~~~
{\sc D}^s_{\sigma}(\mi  p_{\z}; t) = \mathcal{G}^{\mi \1}(\theta; t)\,{\sc D}^s(\mi  p_{\z},m_{\i})\,\mathcal{G}(\theta; t).
\nonumber
\end{equation}
By following the BV prescription
\cite{Bla95}, which takes into account the above definitions,
it was demonstrated \cite{Bla98} that the flavor conversion
formula can be written as
\begin{eqnarray}
P(\mbox{\boldmath$\nu_\alpha$}\rightarrow\mbox{\boldmath$\nu_\beta$};t)
            &=& \left|\left\{{\sc B}^s_{\beta}( p_{\0}; t),\,{\sc B}^s_{\alpha}( p_{\0}; t)\right\}\right|^{\2}
            + \left|\left\{{\sc D}^s_{\beta}(\mi p_{\0}; t), \,{\sc B}^s_{\alpha}( p_{\0}; t)    \right\}\right|^{\2}
\label{A25}
\end{eqnarray}
which is calculated without considering the
localization conditions imposed by wave packets, i. e. by assuming that $ p_{\z} \approx  p_{\0}$.
When the explicit form of the flavor annihilation and creation
operators are substituted in Eq.~(\ref{A25}), it was also demonstrated \cite{Bla03B} that
the flavor oscillation formula becomes
\begin{eqnarray}
P(\mbox{\boldmath$\nu_\alpha$}\rightarrow\mbox{\boldmath$\nu_\beta$};t)
            &=& \mbox{$\frac{\sin^{\2}{[2\theta]}}{2}$}\left\{\left[1 - f( p_{\0},m_{\1},m_{\2}) \, \right] \,
            \cos [ \epsilon_{\mi}( p_{\0},m_{\1},m_{\2}) \, t ]
            \right. \nonumber\\
            &&\left.~~~~~~~~~~~~~~~~~~~~~~~
            + f( p_{\0},m_{\1},m_{\2}) \, \cos [
\epsilon_{\pl}( p_{\0},m_{\1},m_{\2}) \, t] \right\}\nonumber\\
            &\approx& \mbox{$\sin^{\2}{[2\theta]}
            \left\{\left[1 - \left(\frac{\Delta m}{2 p_{\0}}\right)^{\2}\right]\sin^{\2}{\left[\frac{\Delta m^{\2}}{4  p_{\0}}t\right]}
            + \left(\frac{\Delta m}{2 p_{\0}}\right)^{\2} \sin^{\2}{\left[ p_{\0}t\left(1 + \frac{m_{\1}^{\2} + m_{\2}^{\2}}{4  p_{\0}^{\2}}\right)\right]}\right\}~~$}
            \label{A26}
\end{eqnarray}
where the last approximation takes place in the relativistic limit $ p_{\0} \gg \sqrt{m_{\1} m_{\2}}$.
After some simple mathematical manipulations,
the Eq.~(\ref{A26}) gives exactly
the oscillation probability $ P_{\Dirac}(\mbox{\boldmath$\nu_\alpha$}\rightarrow\mbox{\boldmath$\nu_\beta$};L)$
calculated from Eq.~(\ref{A14}) when it is assumed that the wave packet width $a$ tends to infinity and $t \approx L$.

This new oscillation formula tends to the standard one
(\ref{20A}) in the UR limit. If the mass eigenstates
were nearly degenerate, we could have focused on the case of a
nonrelativistic oscillating particle having {\em very} distinct mass
eigenstates. Under these conditions, the quantum theory of
measurement says that interference vanishes.
Therefore, as we have already appointed, the
effects are, under realistic conditions, far from
observable. Besides, in spite of working on a QFT framework, the lack
of observability conditions must be overcome by implementing {\em
external} wave packets, i. e. by calculating the
explicit form of Eq.~(\ref{A17}) for fermions.
Such a procedure was applied by Beuthe for scalar particles \cite{Beu03} and,
in a very particular analysis,
with basis on the BV calculations and on
our {\em intermediate} wave packet results, it could be extended to
the fermionic case.

\section*{IV. Flavor and Chiral Oscillations}

In treating the time evolution of the spinorial
mass-eigenstate wave packets in the previous section,
we have overlooked an important feature.
We have {\em completely} disregarded
the chiral nature of charged weak currents and the time evolution
of the chiral operator. In the following, we aim to investigate
if (and eventually how) the flavor oscillation formula could be
modified by this additional effect.

It is well known that from the Heisenberg equation, we can
immediately determine whether or not a given observable is a
constant of the motion. If neutrinos have mass, the operator $\gamma^{\5}$ does
not commute with the mass-eigenstate Hamiltonians. This means that
for massive neutrinos chirality is {\em not} a constant of the
motion. Observing that neutrinos with positive chirality are
decoupled from charged weak currents, this additional effect
cannot be ignored.
We have already seen that localized states
contains, in general,  plane-wave components of negative and
positive frequencies. As an immediate consequence of this, the
interference between positive and negative frequencies,
responsible for the additional oscillatory term in $\mbox{\sc
Dfo}(t)$, will also imply an oscillation in the average  of
chirality. Thus, the use of Dirac equation as evolution equation
for neutrino mass-eigenstate wave packets leads to an oscillation
formula containing both  ``flavor-appearance'' (neutrinos of a
flavor not present in the original source) and
``chiral-disappearance'' (neutrinos of wrong chirality)
probabilities.

We obtain the Dirac flavor and chiral oscillation probability formula in the
same way as we have obtained the Eq.~(\ref{A14BB}).
By assuming that the
normalizable mass-eigenstate wave functions $\psi_{\1,\2}(z,t)$
are created  at time $t=0$ as a $-1$ chiral eigenstate,
we can write
\begin{eqnarray}
\lefteqn{Re\left\{\int_{_{-\infty}}^{^{+\infty}} dz \,
\psi^{\dagger}_{\i}(z,t) \,  \gamma^{\5} \, \psi_{\j}(z,t)\right\} =
\int_{_{-\infty}}^{^{+\infty}}\frac{d p_{\z}}{2 \pi} \, \varphi^2( p_{\z} - p_{\0})}\nonumber\\
&&~~~~~~~~\times
\left\{ \, \left[ \, 1 -
f( p_{\z},m_{\i},m_{\i})  - \, \frac{m_{\i}
m_{\i}}{E( p_{\z},m_{\i})E( p_{\z},m_{\i})} \, \right] \, \cos [
\epsilon_{\mi}( p_{\z},m_{\i},m_{\i}) \, t ] \right.
\nonumber
\\  & & ~~~~~~~~~~~~~~~~~~ +
\left. \left[ \, f( p_{\z},m_{\i},m_{\i}) +
\frac{m_{\i} m_{\i}}{E( p_{\z},m_{\i})E( p_{\z},m_{\i})} \, \right]
\, \cos [ \epsilon_{\pl}( p_{\z},m_{\i},m_{\i}) \, t] \right\}
 ~~
(\mbox{\footnotesize $i,j \,= \,1,2$})~.
\end{eqnarray}
From this integral, it is readily seen that an initial $-1$ chiral
mass-eigenstate will evolve with time changing its chirality. Once
we know the time evolution of the chiral operator, we are able to
construct an {\em effective} oscillation probability which takes
into account both flavor and chiral conversion effects, i.e.
\begin{eqnarray}
P(\mbox{\boldmath$\nu_{\alpha,\L}$}\rightarrow\mbox{\boldmath$\nu_{\beta,\L}$};t)
& = &
\int_{_{-\infty}}^{^{+\infty}} dz \,
|\psi_{\b,\L} (z, t;\theta)|^{\2}
= \int_{_{-\infty}}^{^{+\infty}} dz\,
\mbox{$\psi^{\prime}_{\b} (z, t;\theta) \, \frac{1-\gamma^{\5}}{2}
 \, \psi_{\b}(z, t;\theta)$}
\nonumber \\
 & = &\mbox{$\frac{\sin^{\2}{[2\theta]}}{2}$} \, \left\{ \, \frac{1}{2}\,\sum_{i=1}^{2} \,
\left[\int_{_{-\infty}}^{^{+\infty}} dz \,
|\psi_{\i,\L} (z, t)|^{\2} \right]-
Re \left[\int_{_{-\infty}}^{^{+\infty}} dz \,
\psi_{\1,\L}^{\prime} (z, t) \,
\psi_{\2,\L}(z, t) \right] \right\} \nonumber \\
& = & \mbox{$\frac{\sin^{\2}{[2\theta]}}{2}$}
\left[\mbox{\sc Dco}(t) - \mbox{\sc Dfco}(t)\right]
.
\label{41}
\end{eqnarray}
As done in the previous section, the terms $\mbox{\sc Dco}(t)$ and
$\mbox{\sc Dfco}(t)$ can be rewritten by using a
$ p_{\z}$-integration,
\begin{eqnarray}
\mbox{\sc Dco}(t) & = & \frac{1}{2} \sum_{i=1}^{2}\,
\int_{_{-\infty}}^{^{+\infty}} \frac{dp_z}{2 \, \pi}\,
\varphi^{\2}(p_{\z} - p_{\0}) \, \left\{ \, 1 - c( p_{\z},m_{\i},m_{\i}) +
c( p_{\z},m_{\i},m_{\i}) \, \cos [2 \,
E( p_{\z},m_{\i}) \, t] \right\} \nonumber \\
 & =& 1 -  \frac{1}{2} \sum_{i=1}^{2} \, \int_{_{-\infty}}^{^{+\infty}}
\frac{dp_z}{2 \, \pi}\,
\varphi^{\2}(p_{\z} - p_{\0}) \, \left\{ \,
\frac{m_{\i}^{\2}}{2 \, E^{^{\, \2}}( p_{\z},m_{\i})} -
\frac{m_{\i}^{\2}}{2 \, E^{^{\, \2}}( p_{\z},m_{\i})} \cos [ 2 \,
E( p_{\z},m_{\i}) \, t] \right\}
\label{41A}\end{eqnarray}
and
\begin{eqnarray} \mbox{\sc Dfco}(t) & = &
\int_{_{-\infty}}^{^{+\infty}}
\frac{dp_z}{2 \, \pi}\,
\varphi^{\2}(p_{\z} - p_{\0})
 \, \left\{ \, \left[ \, 1 -
c( p_{\z},m_{\1},m_{\2}) \, \right] \, \cos [
\epsilon_{\mi}( p_{\z},m_{\1},m_{\2}) \, t ] \right.  \nonumber
\\  & & ~~~~~~~~~~~~~~~~~~~~~~~~~~~~~~~~~~~ ~~~~~~+ \left. c( p_{\z},m_{\1},m_{\2}) \, \cos [
\epsilon_{\pl}( p_{\z},m_{\1},m_{\2}) \, t] \right\} ~,
\label{41B}\end{eqnarray}
where
\[
c( p_{\z},m_{\i},m_{\j}) =  f( p_{\z},m_{\i},m_{\j}) + \frac{m_{\i}
m_{\j}}{2 \, E( p_{\z},m_{\i})E( p_{\z},m_{\j})}~.
\]
The functions $c( p_{\z},m_{\i},m_{\j})$ have a common maximum at
$ p_{\z}=0$ which, contrary to what happened for
$f( p_{\z},m_{\1},m_{\2})$, do not depend on the mass values,
\[ c_{\mbox{\tiny max}}(0,m_{\i},m_{\j})=\mbox{$\frac{1}{2}$}\, \, ,\]
and, following the same asymptotic behavior of
$f( p_{\z},m_{\1},m_{\2})$, go rapidly to zero for $ p_{\z}\gg
m_{\1,\2}$.
As a consequence of the first order approximation (\ref{34AA}),
we get
\begin{eqnarray*}
c( p_{\z},m_{\i},m_{\j}) & \approx & \mbox{$\frac{1}{2}$} \, \left[
1 - \mbox{v}_{\i}\mbox{v}_{\j} + \mbox{v}_{\i}\mbox{v}_{\j} \, \left(\mbox{v}_{\i}^{\2} + \mbox{v}_{\j}^{\2}
-2 \right) \, \frac{ p_{\z}- p_{\0} }{ p_{\0}} \, \right].
\end{eqnarray*}
which gives
\begin{eqnarray*}
c( p_{\z},m_{\1},m_{\2}) & \approx & \frac{m_{\1}^{\2}+
m_{\2}^{\2}}{4\, p_{\0}^{\, \2}} \, \left( \, 1 -  2 \, \,
\frac{ p_{\z}- p_{\0} }{ p_{\0}} \, \right)
\end{eqnarray*}
in the UR approximation.
By substituting $c( p_{\z},m_{\i},m_{\j})$ in the above integrations (\ref{41A}-\ref{41B}) and
after some algebraic manipulations, we
explicitly calculate the terms $\mbox{\sc
Dco}(t)$ and $\mbox{\sc Dfco}(t)$,
\begin{eqnarray}
\mbox{\sc Dco}(t) & \approx & \mbox{$1 -
 \frac{m_{\1}^{\2}}{4\, p_{\0}^{\, \2}}  +
 \exp \left[ - \left( \, \frac{ 2 \, p_{\0}^{\, \2} - m_{\1}^{\2}}{\sqrt{2} \,a \,
  p_{\0}^{\, \2}} \, t \right)^{\2} \, \right] \,
 \frac{m_{\1}^{\2}}{4\, p_{\0}^{\, \2}}
 \,  $}\nonumber \\
& & \mbox{$\left.  \times \left\{ \cos \left[ \frac{2 \, p_{\0}^{\, \2} +
m_{\1}^{\2}}{ p_{\0}} \, t \right] + \frac{4 \, p_{\0}^{\, \2} - 2
\, m_{\1}^{\2}}{a^{\2}  p_{\0}^{\, \3}} \, t \, \sin \left[ \frac{2
\, p_{\0}^{\, \2} + m_{\1}^{\2}}{ p_{\0}} \, t \right]\, \right\} \,
\right\} \, $}\nonumber \\
&  &\mbox{$ -
 \frac{m_{\2}^{\2}}{4\, p_{\0}^{\, \2}}  +
 \exp \left[ - \left( \, \frac{ 2 \, p_{\0}^{\, \2} - m_{\2}^{\2}}{\sqrt{2} \,a \,
  p_{\0}^{\, \2}} \, t \right)^{\2} \, \right] \,
 \frac{m_{\2}^{\2}}{4\, p_{\0}^{\, \2}} \, $}\nonumber
\\
 & &\mbox{$ \times \left.  \left\{ \,  \cos \left[ \frac{2 \, p_{\0}^{\, \2} +
m_{\2}^{\2}}{ p_{\0}} \, t \right] + \frac{4 \, p_{\0}^{\, \2} - 2
\, m_{\2}^{\2}}{a^{\2}  p_{\0}^{\, \3}} \, t \, \sin \left[ \frac{2
\, p_{\0}^{\, \2} + m_{\2}^{\2}}{ p_{\0}} \, t \right]\, \right\} \,
\right\} \, \,$} ,\\
 \mbox{\sc Dfco}(t) & \approx & \mbox{$\exp \left[ -
\left( \, \frac{ \Delta m^{\2} }{2\sqrt{2} \,a \,  p_{\0}^{\, \2}}
\, t \right)^{\2}
\right] \,$}\nonumber \\
 &  &  \mbox{$ \times \left\{ \,
\, \left[ 1 - \, \frac{m_{\1}^{\2} + m_{\2}^{\2}}{4\, p_{\0}^{\,
\2}} \, \right] \,
 \cos \left[ \frac{\Delta m^{\2}}{2\, p_{\0}} \, t \right]
 + \frac{m_{\1}^{\2} + m_{\2}^{\2}}{4\, p_{\0}^{\,
\2}}
 \, \frac{\Delta m^{\2} }{a^{\2}  p_{\0}^{\, \3}} \, \, t \, \sin
\left[ \frac{\Delta m^{\2}}{2\, p_{\0}} \, t \right]
\,\right\}$} \nonumber \\
 & & \mbox{$ + \exp \left[ - \left( \, \frac{ 4 \, p_{\0}^{\, \2} - m_{\1}^{\2} -
m_{\2}^{\2}}{2\sqrt{2} \,a \,  p_{\0}^{\, \2}} \, t \right)^{\2} \,
\right] \, \frac{m_{\1}^{\2} + m_{\2}^{\2}}{4\, p_{\0}^{\, \2}}
\, $} \nonumber \\
 & & \mbox{$\left\{ \,
\times \cos \left[ \frac{4 \, p_{\0}^{\, \2} + m_{\1}^{\2} +
m_{\2}^{\2}}{2 \,  p_{\0}} \, t \right] +  \frac{4 \, p_{\0}^{\, \2}
- m_{\1}^{\2} - m_{\2}^{\2}}{a^{\2}  p_{\0}^{\, \3}} \, t \, \sin
\left[ \frac{4 \, p_{\0}^{\, \2} + m_{\1}^{\2} + m_{\2}^{\2}}{2\,
 p_{\0}} \, t \right]\, \right\}  \, \,$} .
\end{eqnarray}
Again, in
the hypothesis of minimal slippage between the mass-eigenstate
wave packets ($\Delta v L\ll a$), and for long distance between
source and detector  ($L \gg a$),i.e.
\[
1 \, \ll \frac{L}{a} \ll \frac{ p_{\0}^{\, \2}}{\Delta \, m^{\2}}
\, , \,
\]
the standard flavor oscillation probability is reproduced. In
fact,
\begin{eqnarray}
 P \left( \boldsymbol{\nu_{\a,\L}}  \to \boldsymbol{\nu_{\b,\L}}
; \, L  \right)
& \approx & \mbox{$\frac{\sin^{\2}{[2\theta]}}{2}$}
\, \mbox{$\left[ 1 - \, \frac{m_{\1}^{\2} +
m_{\2}^{\2}}{4\, p_{\0}^{\, \2}} \, \right]
\left\{ \, 1 - \left[ 1 - \left( \, \frac{ \Delta m^{\2}
}{2\sqrt{2} \,a \,  p_{\0}^{\, \2}} \, L \right)^{\2} \right] \,
\cos \left[ \frac{\Delta m^{\2}}{2\, p_{\0}} \, L \right] \,
\right\}$}  \nonumber \\
& \approx &\mbox{$\frac{\sin^{\2}{[2\theta]}}{2}$}
\,\mbox{$ \left\{ \, 1 - \cos \left[ \frac{\Delta
m^{\2}}{2\, p_{\0}} \, L \right] \,
\right\}$}  \nonumber \\
 & = & \sin^{\2}[
2 \theta] \, \, \,\mbox{$\sin^{2} \left[ \frac{\Delta m^{2}}{4  p_{\0}} \,
L \right]$}~.
\end{eqnarray}

\section*{V. Conclusions}

In order to quantify some subtle changes which appear in the standard flavor oscillation probability \cite{Kay04}
due to chiral oscillations coupled to the flavor conversion
mechanism of free propagating wave packets, we have reported about
some recent results on the study of flavor oscillation with Dirac wave packets \cite{Ber04}.
By taking into account the spinorial
form of neutrino wave functions and imposing an initial constraint
where a {\em pure} flavor-eigenstate is created at $t = 0$,
for a constant spinor $w$,
it is possible to calculate the contribution of positive and negative frequency solutions
of the Dirac equation to the wave packet propagation and, finally, to obtain the oscillation probability.
Particularly, we have noticed a term of very high oscillation frequency depending on the sum of energies
in the new oscillation probability formula.
In addition, the spinorial form of the wave functions and their
chiral oscillating character subtly modify
the coefficients of the oscillating terms in this flavor conversion formula.
To describe the time evolution of the mass-eigenstates, we have
assumed an initial {\em gaussian}
localization and performed integrations by considering a strictly
peaked momentum distribution.
Under the particular assumption of UR particles,
we have been able to obtain an
analytic expression for the coupled chiral and flavor conversion formula.
In case of Dirac wave packets, these modifications introduce correction factors which are negligible in the UR limit.
We have confirmed that the {\em fermionic} character of the particles modify the standard oscillation
probability which is previously obtained by implicitly assuming a {\em scalar} nature of
the mass-eigenstates.

However, we know the necessity of a more sophisticated approach
is understood. It involves a field-theoretical treatment.
Derivations of the oscillation formula resorting to field-theoretical methods are not very
popular. They are thought to be very complicated and the existing quantum field computations of the
oscillation formula do not agree in all respects \cite{Beu03}.
The  Blasone and Vitiello (BV) model \cite{Bla95,Bla03} to neutrino/particle mixing and oscillations
seems to be the most distinguished trying to this aim.
They have attempt to define a Fock space of weak eigenstates and to derive a nonperturbative oscillation formula.
Flavor creation and annihilation operators, satisfying canonical (anti)comutation relations, are defined by means of Bogoliubov transformations.
As a result, new oscillation formulas are obtained for fermions and bosons, with the oscillation frequency depending not only on the difference but also on the sum of the energies of the different mass-eigenstates.
Meanwhile, the prescription of oscillating neutrinos as Dirac spinors was
not yet completely and accurately described in a quantum field formalism.
With Dirac wave packets, the flavor conversion formula can be reproduced \cite{Ber04B}
with the same mathematical structure as those obtained in the BV model \cite{Bla95,Bla03}.
Moreover each new effect present in the oscillation formula can be separately quantified.

In fact, the quantum-mechanical treatment which associates Dirac
wave packets with the propagating mass eigenstates is rich in physical insights
which were extensively studied in this paper.
Besides the review of analytical calculations done with {\em gaussian}
wave packets for {\em scalar} \cite{Ber04} and {\em fermionic} \cite{Ber04B}
particles, the main conceptual aspect arising from the formalism with
Dirac wave packets leads to the study of chiral oscillations.
In the standard model flavor-changing interactions, neutrinos with positive
chirality are decoupled from the neutrino absorbing charged weak currents \cite{DeL98}.
A state with {\em left-handed} helicity can be approximated by
a state with negative chirality in the UR limit.
Once we have assumed the interactions at the source and detector are chiral
only the component with negative chirality contributes to the propagation.
In this case, we are obliged to consider chiral coupled to flavor oscillations
in order to compute the modifications to the standard flavor conversion formula.
In fact, when chiral oscillations are taken into account,
these modifications introduce correction factors
proportional to $m_{\1,\2}^{\2}/p_{\0}^{\2}$ which are, however, practically
un-detectable by the current experimental analysis.
It leads to the conclusion that, in spite of often being criticized,
the standard flavor oscillation formula resorting to the plane wave derivation
can be reconsidered when {\em properly} interpreted,
but a satisfactory description
of {\em fermionic} (spin one-half) particles requires the use of the
Dirac equation as evolution equation for the mass-eigenstates.

\section*{Acknowledgments}

The authors thank the University of Lecce for the hospitality
and the CAPES (A.E.B.) and FAEP (S.D.L.) for
Financial Support.

\newpage

\newpage

\section*{Figures}

\end{document}